\documentclass[usenatbib]{mn2e}

\usepackage{amssymb,amsmath}
\usepackage{graphicx}
\usepackage{multirow}
\usepackage{natbib}
\usepackage{lineno}
\def\citeN{\citet}
\def\cite{\citep}

\footnotesize
\newdimen\digitwidth    
\setbox0=\hbox{\rm0}
\digitwidth=\wd0
\catcode`!=\active
\def!{\kern\digitwidth}
\normalsize
\title[17 and 24 GHz observations of southern pulsars]{17 and 24 GHz observations of southern pulsars}
\author[M.~J.~Keith et al.]
{M.~J.~Keith$^{1}$\thanks{Email: mkeith@pulsarastronomy.net},
S.~Johnston$^{1}$,
L.~Levin$^{2,1}$
M.~Bailes$^{2,3}$
\\
$^1$ Australia Telescope National Facility, CSIRO Astronomy \& Space Science, P.O. Box 76, Epping, NSW 1710, Australia\\
$^2$ Swinburne University of Technology, Centre for Astrophysics and Supercomputing Mail H39, PO Box 218, VIC 3122, Australia\\
$^3$ University of California, Berkley, 601 Campbell Hall 3411, Berkley, CA 94720, USA\\
}

\let\leqslant=\leq 
\date{}
\begin{document}

\maketitle
\newcommand{\setthebls}{
}

\setthebls

\begin{abstract} 
We present observations of PSRs J0437--4715, J0738--4042, J0835--4510, J0908--4913, J1048--5832 , J1622--4950, J1644--4559, J1721--3532 and J1740--3015 at 17~GHz using the Parkes radio telescope.
All 9 were detected at 17 GHz, additionally, we detected PSR J0835--4510 and J1622--4950 at 24 GHz.
Polarisation profiles of each pulsar and the variation with frequency are discussed.
In general, we find that the highly polarised edge components of young pulsars continue to dominate their profiles at 17~GHz.
Older pulsars ($\gtrsim 10^{5}$ years) appear to be almost completely depolarised.
Our detection of PSR J0437--4715 is the highest frequency observation of a millisecond pulsar to date, and implies a luminosity at 17~GHz of 14~$\mu$Jy$\,$kpc$^2$, and a mean spectral index of $2.2$.

We find that the spectral index of the magnetar PSR J1622--4950 is flat between 1.4 and 24~GHz, similar to the other known radio magnetars XTE J1810--197 and 1E 1547.0--5408.
The profile is similar to that at 3.1~GHz, and is highly linearly polarised.
Analysis of the frequency evolution of the profile of PSR J0835--4510 show that the profile is made of four components that vary with frequency only in their amplitude.
The width and separation of the components remains fixed and the spectral index of each component can be determined independently.

\end{abstract}


\section{Introduction}
Radio pulsar emission is a broad band phenomenon, detected from 10~MHz to 150~GHz, however since the flux density typically drops off very rapidly with frequency, most observations are carried out at frequencies below 4~GHz.
Indeed, the majority of pulsars have spectra which can be modelled by a power law, $S\propto \nu^{-\alpha}$, with a mean spectral index, $\alpha = 1.8\pm0.2$ \cite{mkkw00}.
With the current generation of radio telescopes, higher frequency observations are only viable for the most luminous pulsars, or those with an unusually flat spectral index.
Probing a wide range of frequencies is one key to understanding the nature of the emission processes in the pulsar magnetosphere.

Pulsars exhibit a wide variety of profile shapes, however one can typically break them up into a number of distinct or overlapping components which are typically roughly Gaussian in shape.
In order to begin to interpret the observed profiles, we classify them, typically into so-called `core' and `conal' components \cite{kmc70,bac76,ran83}.
Core components are thought to be emitted close to the magnetic axis and typically have a steep spectral index compared to the conal components that are thought to arise from the edges of the emission region.

Other than the flux density, there are a myriad of other frequency dependent effects observed in pulsars.
Most prominently is the negative-index power-law relationship between observing frequency and overall pulse width noted in many pulsars \cite{tho91a}.
This is most simply understood by a mapping between the emission height (and therefore the spread angle of the magnetic field lines) and the plasma frequency and therefore the frequency of the emitted radio waves \cite{rs75}.
This effect is most evident at frequencies below $\sim 1$~GHz, and at higher frequencies there does not seem to be a significant change of pulse width with frequency.
There is also evidence that this effect seems to be constrained to the outer `conal' components, and that inner cones and core emission do not share this frequency dependent width effects \cite{mr02a}.
\citeN{gg03} suggest that the emission height of components is longitude dependent, and that the outer components are emitted further from the neutron star surface.
Indeed, it has been suggested that the observed pulse shape may be entirely caused by mixing components from many emission heights at a single frequency, rather than the traditional model of radially separated components \cite{kj07}.

Pulsar radio emission is often highly polarised, and the polarisation properties observed to be frequency dependent.
In general, polarisation fraction decreases with frequency, however \citeN{hkk98} showed that there are pulsars for which a highly polarised component of the profile with a relatively flat spectral index becomes prominent at higher frequencies.
\citeN{jkw06} observed 32 southern radio pulsars at a frequency of 8.3~GHz, and demonstrated that whilst most pulsars are depolarised with increasing frequency, many young pulsars exhibit such polarised flat spectrum components and were typically classified as partial cones.
\citeN{jw06} suggest that the emission of young pulsars is highly polarised conal emission, and arises from high in the magnetosphere.

At higher frequencies still, \citeN{wjkg93} and \citeN{kjdw97} report detections of PSRs B0329+54, B0355+54, B1929+10 and B2021+51 at 34 and 43~GHz.
The strength of the detection at these frequencies suggest that the spectra of pulsars may flatten, or indeed turns upwards at millimetre wavelengths.

The highest frequency for which a radio detection of a pulsar has been published is observations of the radio magnetar XTE J1810--197.
This, the first known radio magnetar, was identified by \citeN{crj+07} by radio emission associated with an outburst of the X-ray pulsar.
The pulsed radio emission was shown to be highly polarised, and with a spectral index close to 0 and detectable as high as 144 GHz \cite{crp+07}.
A second magnetar, 1E 1547.0--5408, was also found to exhibit radio emission with a flat spectral index and high degree of polarisation\cite{crhr07,crj+08}.
A third member of this class, PSR J1622--4950, has now been identified by \citeN{lbb+10}.
This object was discovered as part the High Time Resolution Universe survey for radio pulsars \cite{kjv+10}.
Although there is, as yet, no evidence for a magnetar-like X-ray outburst in PSR J1622--4950, the magnetic field strength inferred from the rotational properties, $3 \times 10^{14}$~G, is the highest known of any radio pulsar and similar to other magnetars.
Indeed, the high polarisation, flat spectrum and large pulse width observed in PSR J1622--4950 are properties shared with the other two known radio magnetars.

Here we present recent observations of PSR J1622--4950 and 8 other southern radio pulsars at 17 and 24 GHz, the highest frequency that can be achieved at the Parkes radio telescope.
These were selected based on their 8.3~GHz flux density \cite{jkw06}.
In the next section we outline the observational setup and data analysis procedure and in section 3 we present the results and discuss the implications of our observations.

\section{Observations and Analysis}
Observations were carried out on 2010 November 1st and 2nd at the Parkes radio telescope.
We used the `13-mm' receiver and the Parkes Digital Filterbank System `PDFB3' at centre frequencies of 17.0~GHz and 24.0~GHz.
Each of the pulsars were observed for 1 hour with a 1024~MHz bandwidth centred at 17.0~GHz.
For J0835--4510 and J1622--4950 we were able to get sufficient detection significance at 17~GHz to suggest that observation at 24~GHz was warranted.
These two pulsars were observed with a 1024~MHz bandwidth centred at 24.0~GHz.
Additionally, the observation of PSR J1622--4950 was repeated several times with a total of 5.4 hours at 17~GHz and 2.4 hours at 24~GHz.

Calibration and analysis of the data was carried out using the {\sc psrchive} software suite \cite{hvm04}.
Before the start of each observation we used a noise diode in the receiver to calibrate for differential gain and leakage between the linear feeds.
It should be noted that the observed position angles have an arbitrary zero point.
Flux calibration was performed by measuring the strength of the noise diode to the standard calibrator PKS 1253--055.
We assume a flux density of 20~Jy for PKS 1253--055 at both 17 and 24~GHz, and inferred a typical system equivalent flux density of $\sim 240$~Jy and $\sim 560$~Jy at 17 and 24~GHz respectively.
Fluxes were corrected for elevation effects using a standard gain-elevation curve\footnote{Provided by Parkes operations staff, determined empirically from observations 22~GHz observations of a known source.} parametrised as:
\begin{equation}
G = -0.19403 + 0.045772 \theta - 4.3866 \times 10^{-4} \theta^2,
\end{equation}
where $\theta$ is the mean elevation of the telescope at the time of the observation.
The dominant error in our flux measurements is likely to be system temperature variations during observations of the flux calibrator, and variations in the level of atmospheric absorption.
Repeated observations of the flux calibrator lead us to believe that the flux estimates have a systematic error of $\sim 20\%$. 

Due to the long pulse period of PSR J1622--4950, variations in the system temperature are on the same timescale as the pulse period.
To correct for the unstable time-baselines across the pulse period we fit and subtracted a linear slope from the intensity in each 30 second integration.
No other pulsars were affected by this due to their shorter pulse periods.

\section{Results}
We successfully detected PSRs J0437--4715, J0738--4042, J0835--4510, J0908--4913, J1048--5832 , J1622--4950, J1644--4559, J1721--3532 and J1740--3015 at 17~GHz.
Additionally we detected PSRs J0835--4510 and J1622--4950 at 24~GHz.
The polarisation profile of each pulsar is presented and discussed below.

\begin{table*}
\caption{\label{results_table}
The names and properties of the 9 pulsars observed in this work, and their observed parameters at 17 and 24~GHz.
The first three columns contain the name, spin period and characteristic age of the pulsars.
The remaining columns list the properties observed at the observing frequency specified in the fourth column, namely mean flux density, apparent spectral index when compared to 1.4~GHz, linear polarisation fraction, circular polarisation fraction (absolute value and stokes V) and width at $10\%$ and $50\%$ of the peak.
Note that the flux density measurements are likely to be subject to large systematic errors as described in the text.
Polarisation fractions are averaged over the entire profile.
}
\begin{tabular}{lllllllllll}
Name      & $P$ (ms) & $\tau_c$ (Myr) &  Freq (MHz)  &  Flux (mJy) & $\alpha_{1.4}$  &  L/I  &  $|{\rm V}|$/I & V/I    & $W_{10}$ ($^\circ$) & $W_{50}$ ($^\circ$) \\
\hline
J0437--4715 & 5.757   & 6500 & 17000        &  0.58  & 2.2 & $<0.2$ & $<0.2$  & --      & 12.6 & 7.7  \vspace{1mm}\\
J0738--4042 & 374.9   & 3.7  & 17000        &  0.23  & 2.4 & $<0.3$ & $<0.3$  & --      & --   & 17.5 \vspace{1mm}\\
J0835--4510 & 89.35   & 0.011& 17000        &  5.0   & 2.2 & 0.80   & 0.33    & -0.32   & 14.4 & 7.4  \\
            &         &      & 24000        &  3.4   & 2.1 & 0.71   & 0.39    & -0.37   & 15.1 & 7.7  \vspace{1mm}\\
J0908--4913 & 106.8   & 0.11 & 17000        &  0.26  & 1.5 & 0.46   & $<0.3$  & --      & 7.0  & 3.2  \vspace{1mm}\\
J1048--5832 & 123.7   & 0.020& 17000        &  0.36  & 1.2 & 0.5 & $<0.4$  & --      & --   & 11.3 \vspace{1mm}\\
J1622--4950 & 4326    & 0.004& 17000        &  8.0   & -0.2 & 0.80   & 0.12    & -0.10   & 144  & 19.7 \\
            &         &      & 24000        &  5.6   & -0.1 & 0.34   & 0.20    & -0.10   & 156  & 16.8 \vspace{1mm}\\
J1644--4559 & 455.1   & 0.36 & 17000        &  0.16  & 3.0 & $<0.4$ & $<0.4$  & --      & --   & 18.9 \vspace{1mm}\\
J1721--3532 & 280.4   & 0.18 & 17000        &  1.1   & 1.0 & $<0.2$ & $<0.2$  & --      & 16.9 & 7.4  \vspace{1mm}\\
J1740--3015 & 606.9 & 0.021& 17000          &  0.34  & 1.2 & 0.66   & 0.52    & -0.51   & 3.9 &   2.1 \vspace{1mm}\\
\hline

\end{tabular}

\end{table*}

\subsection{PSR J0437--4715}
The only millisecond pulsar in our sample, PSR J0437--4715, has the the 4th highest 1.4~GHz flux density of any known pulsar, and is by far the brightest millisecond pulsar known.
The distance to PSR J0437--4715 is very well constrained to be $0.157 \pm 0.002$ kpc \cite{vbv+08}, and so we determine the 17~GHz pseudo-luminosity, $L'=S_{\rm 17 GHz}\,d^2$ to be $14$~$\mu$Jy$\,$kpc$^2$.

\begin{figure}
\includegraphics[width=8cm]{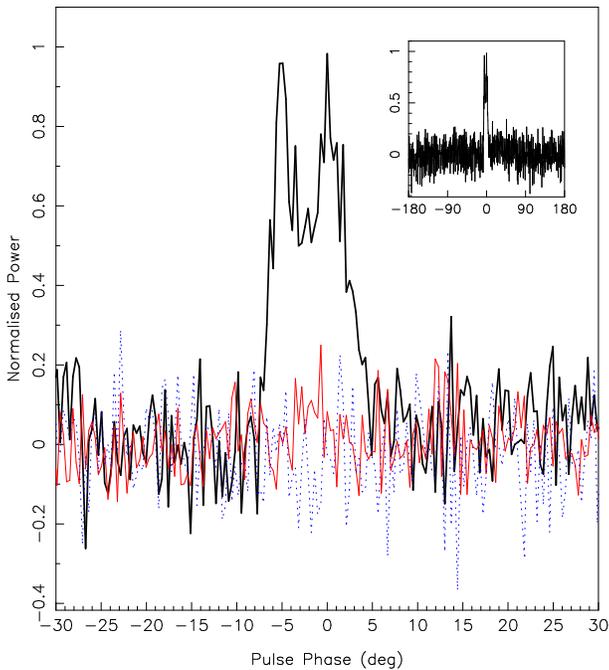}
\caption{
	\label{J0437-4715_17000}
	The pulse profile of PSR J0437--4715 at a centre frequency of 17.0~GHz, showing only phases within 30$^\circ$ of the peak.
		The thick black line shows total intensity, the thin line shows linear polarisation and the dotted line shows circular polarisation.
		Inset is the entire profile over 360 degrees.
}
\end{figure}

\begin{figure}
\includegraphics[width=8cm]{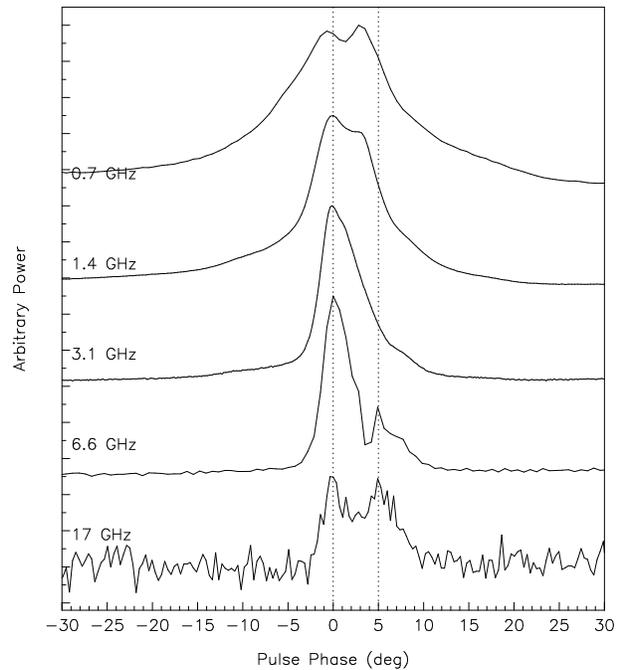}
\caption{
	\label{J0437-4715_mf}
The pulse profile of PSR J0437--4715 in total intensity from 0.7~GHz to 17~GHz, showing only phases within $30^\circ$ of the peak. Data at frequencies below 17~GHz are taken from the Parkes data archive (Hobbs et al., in prep).
The dotted lines indicate the phases of the two peaks visible at 17~GHz.
The profiles are aligned based on the first of the bright peaks in the profile.
}
\end{figure}

At 17~GHz we observe a symmetric, double peaked profile with a bridge of emission.
The polarisation fraction is low, with just a hint of linear polarisation in the centre of the profile.
Figure \ref{J0437-4715_mf} shows the frequency dependent profile evolution of PSR J0437--4715.
Here we present profiles from this work, and from archival Parkes data centred at 732, 1369, 3100 and 6380 MHz, and aligned using the leading component marked with a dotted line in Figure \ref{J0437-4715_mf}.
The profile is complex and very well studied \cite{mj95,nms+97,kll+99}, however the innermost part of the profile, the only part discernible at 17~GHz, clearly consists of at least three components. The central component dominates at low frequencies, below $\sim 1$~GHz, and has the steepest spectrum of the three components. The leading component has a somewhat flatter spectrum and is the only part that is easily identified at all frequencies, being most dominant at 3.1 and 6.7~GHz. Finally, the trailing component has the flattest spectrum, but is much weaker and only identifiable in the 6.7 and 17~GHz profiles.
At lower frequencies still, the profile is dominated by the outer wings (extending up to $\pm 100^\circ$ in pulse longitude), further evidence that the profile is made up of components with a wide range of spectral indices.

It remains difficult to associated the various components in this pulsar, and those seen in other MSPs, with the core/cone dichotomy of \citeN{ran83}.
The symmetry of the profile and abrupt orthogonal mode change at the profile peak clearly observed at e.g. 1.4~GHz seem to imply that the main components is close to the magnetic pole, however the spectral behaviour and location of the outer components does not follow convention and the observed variation in position angles at 1.4~GHz are too complex too attempt to fit with any simple model (e.g. \citealp{ymv+11}).

\subsection{PSR J0738--4042}
PSR J0738--4042 is very weak at 17~GHz, and therefore it is difficult to draw any conclusions.
The lack of polarisation and $\sim 17^\circ$ width is similar to that seen at 8.3~GHz.

\begin{figure}
\includegraphics[width=8cm]{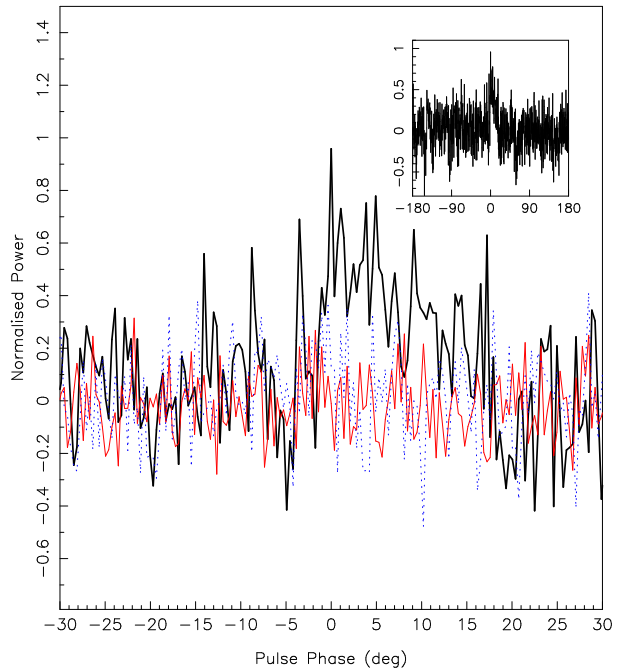}
\caption{
	\label{J0738-4042_17000}
	The pulse profile of PSR J0738--4042 at a centre frequency of 17.0 GHz, showing only phases within 30$^\circ$ of the peak.
	The format is the same as for Figure \ref{J0437-4715_17000}.
}
\end{figure}

\subsection{PSR J0835--4510}
PSR J0835--4510, the Vela pulsar, is easily detected at both 17 and 24~GHz.
At these frequencies we measure flux densities of 5 and 3.4 mJy respectively, however we note that the observed flux density varies somewhat due to interstellar scintillation.
The profile is very similar at both frequencies, symmetrical and roughly Gaussian, with a $10\%$ width of $\sim 15^{\circ}$.
We can see that the polarised fraction is very high, roughly $70\%$ linear and $30\%$ circular polarised at both frequencies.
\citeN{jkw06} observed that the circular fraction increased with frequency between 1.4 and 8.3~GHz, however we find that this trend does not seem to continue, maintaining a $\sim 30\%$ circular fraction out to 24~GHz.

At first glance, the profile at 17~GHz is very similar to the 1.4GHz profile, with pulse width $W_{10} = 14.4^\circ$ (c.f. $W_{10} = 13.3^\circ$ at 1.4~GHz) and strongly polarised.
A closer inspection reveals however that the component seen at 17 and 24~GHz is in fact the weaker trailing component of the low frequency profile.
The frequency evolution of the profile is shown in Figure \ref{J0835-4510_mf}.
Here we present profiles from this work, and from archival Parkes data centred at 1369, 3100, 6380, 8356 and 21840 MHz.
Clearly the spectral index of the leading part of the pulse is much steeper than that of the trailing part.
We investigate the spectral indices of each component in more detail in Section \ref{0835_model_section}.

\begin{figure}
\includegraphics[width=8cm]{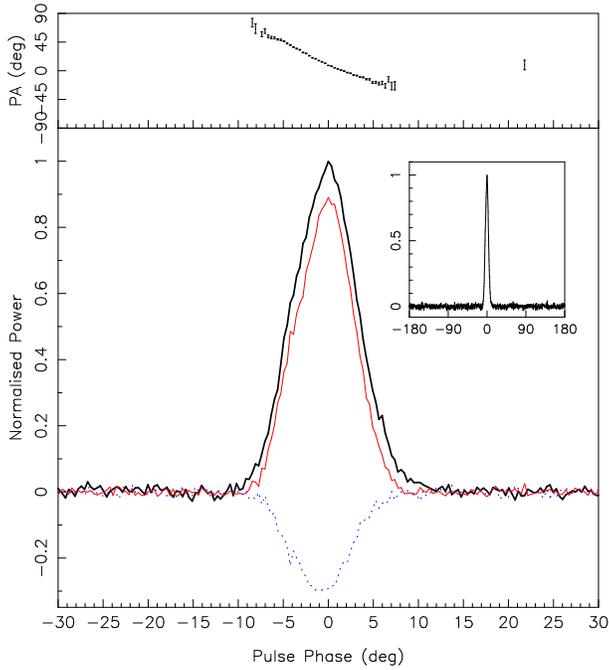}
\caption{
	\label{J0835-4510_17000}
	The pulse profile of PSR J0835--4510 at a centre frequency of 17.0 GHz, showing only phases within 30$^\circ$ of the peak.
		In the lower panel, the thick black line shows total intensity, the thin line shows linear polarisation and the dotted line shows circular polarisation.
		Inset is the entire profile over 360 degrees.
		Above is shown the observed polarisation position angle and its error.
}
\end{figure}

\begin{figure}
\includegraphics[width=8cm]{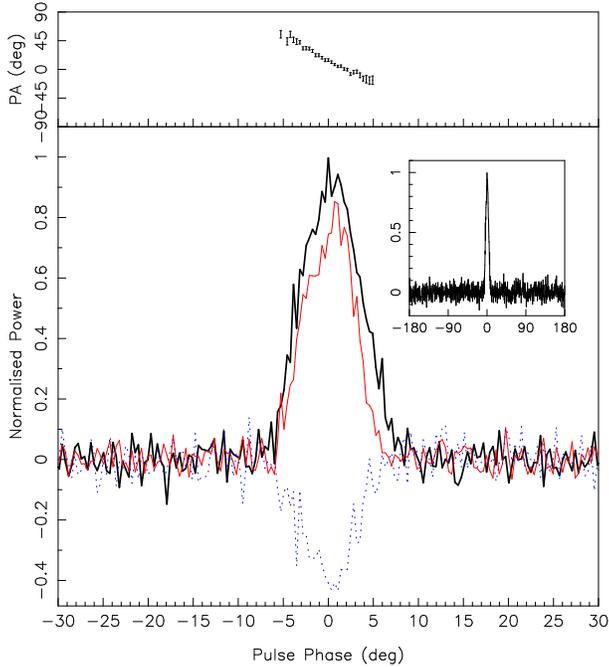}
\caption{
	\label{J0835-4510_24000}
	The pulse profile of PSR J0835--4510 at a centre frequency of 24.0 GHz, showing only phases within 30$^\circ$ of the peak.
		The format is the same as for Figure \ref{J0835-4510_17000}.
}
\end{figure}
\begin{figure}
\includegraphics[width=8cm]{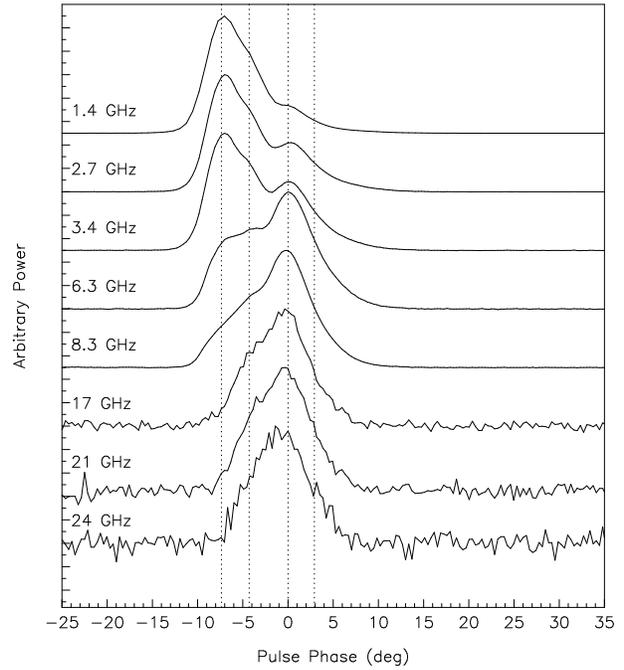}
\caption{
	\label{J0835-4510_mf}
The frequency evolution of the total intensity profile of PSR J0835--4510. The profiles are aligned using the model described in Section \ref{0835_model_section}.
The dotted vertical lines indicate the phase of each of the four components we identify.
}
\end{figure}

\subsection{PSR J0908--4913}
PSR J0908--4913 is a young pulsar with an interpulse approximately 180$^\circ$ from the main pulse and a well defined orthogonal geometry \cite{kj08}.
At 17~GHz we are only able to detect the main pulse, the interpulse is most likely below the detection threshold.
The pulse remains highly polarised, about with a linear and circular polarisation fraction of $50\%$.
This appears to continue the trend observed by \citeN{jkw06}, with the circular fraction increasing with frequency.
The detection significance and pulse width are insufficient to fit model the polarisation position angle swing and further test the observed variation of magnetic inclination angle with frequency \cite{kj08}.

\begin{figure}
\includegraphics[width=8cm]{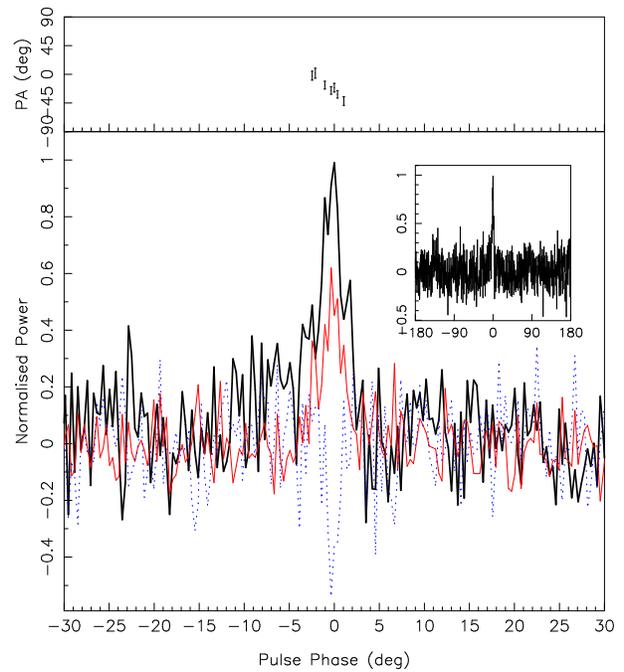}
\caption{
	\label{J0908-4913_17000}
	The pulse profile of PSR J0908--4913 at a centre frequency of 17.0 GHz, showing only phases within 30$^\circ$ of the peak.
		The format is the same as for Figure \ref{J0835-4510_17000}.
}
\end{figure}

\subsection{PSR J1048--5832}

PSR J1048--5832 exhibits a $\sim 4^{\circ}$ pulse width, with a shoulder of about 50\% intensity on the leading edge.
The signal-to-noise is rather low, even in our 1-hour observation, but we infer that this is the polarised central feature observed at lower frequencies, and the shoulder component is the leading component that begins to be apparent at 8.3~GHz \cite{jkw06}.
The data suggests that the leading component has a flatter spectral index than the main component, otherwise it would not be detectable at 17~GHz.

\begin{figure}
\includegraphics[width=8cm]{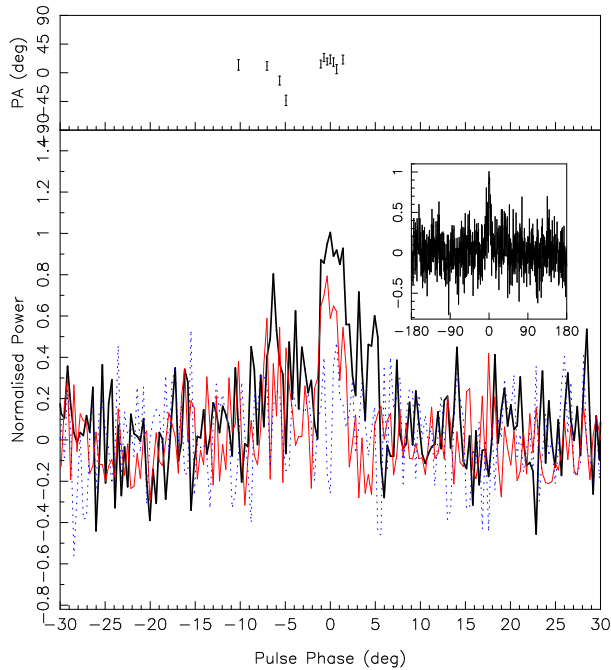}
\caption{
	\label{J1048-5832_17000}
	The pulse profile of PSR J1048--5832 at a centre frequency of 17.0 GHz, showing only phases within 30$^\circ$ of the peak.
		The format is the same as for Figure \ref{J0835-4510_17000}.
}
\end{figure}

\subsection{PSR J1622--4950}
PSR J1622--4950 is the first magnetar discovered in the radio band \cite{lbb+10}.
At 17 and 24~GHz, the profile still covers some $150^\circ$ of pulse phase, and is highly linearly polarised.
It is known that the pulse shape and polarisation is highly variable at lower frequencies, so one must be careful not to infer too much from the high frequency snapshot presented in this paper.
Nevertheless, we observe that the 17~GHz profile is strikingly similar to that most commonly observed at 3.1~GHz, with a high linear fraction, and a small amount of negative circular polarisation at the leading edge of the pulse.
Individual pulses are dominated by narrow, $<5$~ms, impulses which integrate to form the two narrow features at the leading edge of the average profile.
The integrated 24~GHz profile is somewhat more noisy, mainly due to the increased system temperature but appears to show the same general features.
\begin{figure}
\includegraphics[width=8cm]{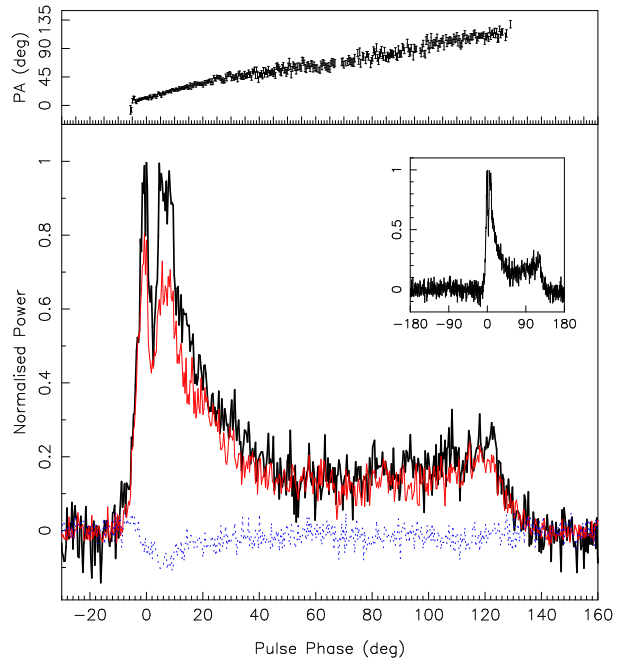}
\caption{
	\label{J1622-4950_17000}
	The pulse profile of PSR J1622--4950 at a centre frequency of 17.0 GHz, showing only phases $-30 < \phi < 160$.
		The format is the same as for Figure \ref{J0835-4510_17000}.
}
\end{figure}

\begin{figure}
\includegraphics[width=8cm]{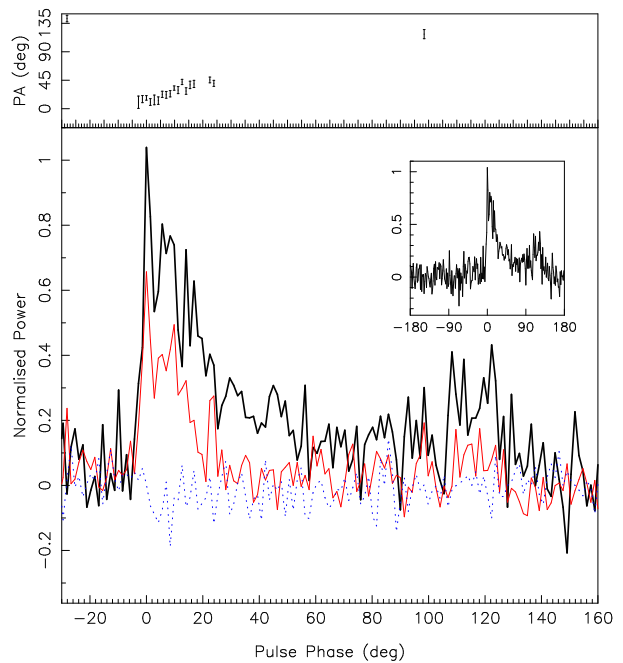}
\caption{
	\label{J1622-4950_24000}
	The pulse profile of PSR J1622--4950 at a centre frequency of 24.0 GHz, showing only phases $-30 < \phi < 160$.
		The format is the same as for Figure \ref{J0835-4510_17000}.
}
\end{figure}
\subsection{PSR J1644--4559}
PSR J1644--4559 is the second brightest pulsar at 1.4~GHz, however at 17~GHz it is barely detectable in our 1 hour observation.
Although the signal-to-noise is low, we note that there appears to be a precursor to the main pulse at a separation of $\sim 10^{\circ}$, which at 17~GHz is nearly as strong as the main component of the pulse.
This is likely to be the same precursor seen as a small shoulder at lower frequencies \cite{jkw06}, and is clearly much flatter in spectral index than the main component of the pulse.
The low signal-to-noise ratio makes it hard to determine the polarisation fraction.

\begin{figure}
\includegraphics[width=8cm]{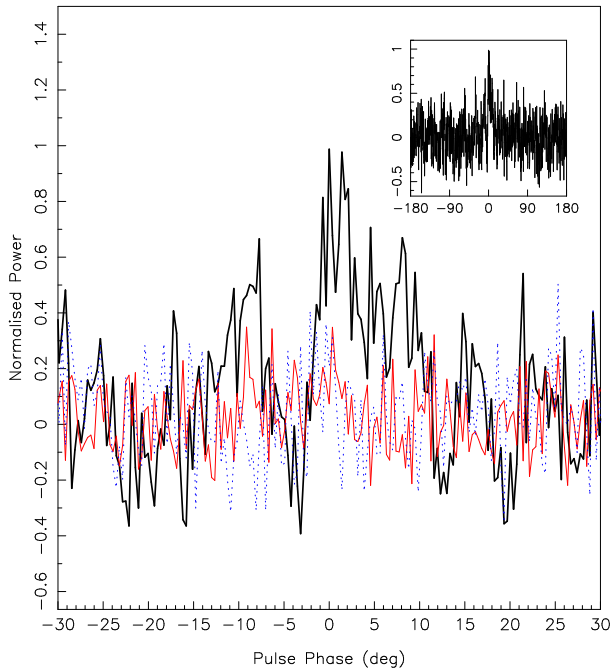}
\caption{
	\label{J1644-4559_17000}
	The pulse profile of PSR J1644--4559 at a centre frequency of 17.0 GHz, showing only phases within 30$^\circ$ of the peak.
		The format is the same as for Figure \ref{J0437-4715_17000}.
}
\end{figure}

\subsection{PSR J1721--5332}
PSR J1721--5332 shows a similar profile at 17~GHz to those at 8.3~GHz and below \cite{jkw06}, with a shallow leading edge and a sharper trailing edge.
At 17~GHz the profile does not have any significant polarisation, and it appears that the polarisation fraction is decreasing with frequency.

\begin{figure}
\includegraphics[width=8cm]{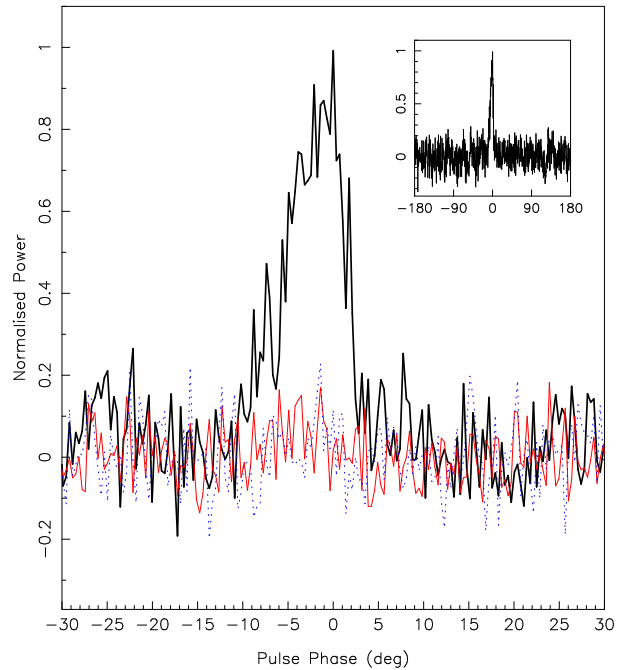}
\caption{
	\label{J1721-3532_17000}
	The pulse profile of PSR J1721--3532 at a centre frequency of 17.0 GHz, showing only phases within 30$^\circ$ of the peak.
		The format is the same as for Figure \ref{J0437-4715_17000}.
}
\end{figure}

\subsection{PSR J1740--3015}
\begin{figure}
\includegraphics[width=8cm]{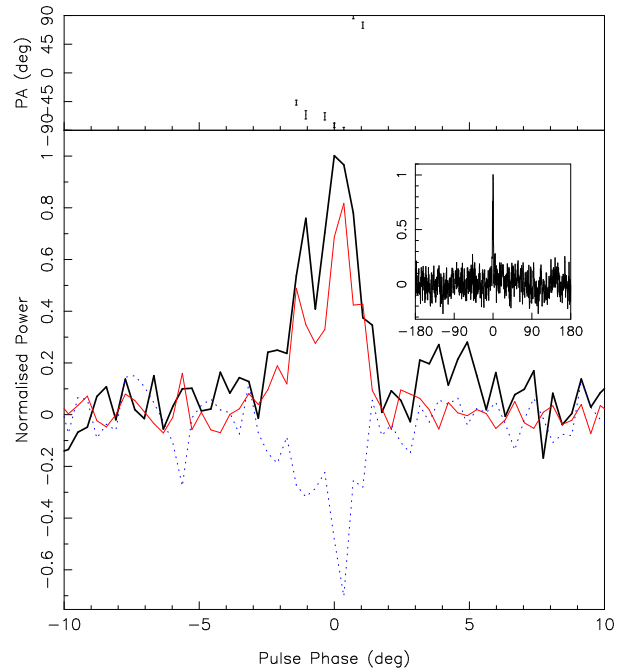}
\caption{
	\label{J1740-3015_17000}
	The pulse profile of PSR J1740--3015 at a centre frequency of 17.0 GHz, showing only phases within 30$^\circ$ of the peak.
		The format is the same as for Figure \ref{J0835-4510_17000}.
}
\end{figure}

\begin{figure}
\includegraphics[width=8cm]{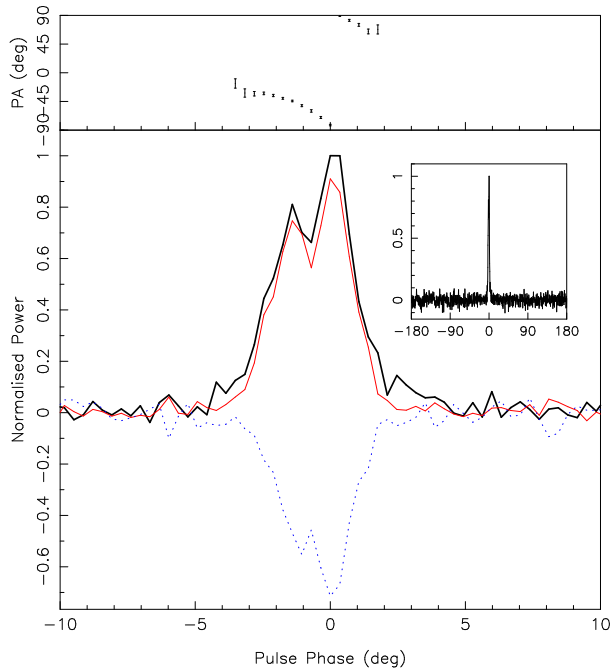}
\caption{
	\label{J1740-3015_8356}
	The pulse profile of PSR J1740--3015 at a centre frequency of 8.356 GHz, showing only phases within 30$^\circ$ of the peak.
This figure is a correction to the profile originally published in \citeN{jkw06} as described in the text.
	The format is the same as for Figure \ref{J0835-4510_17000}.
}
\end{figure}

At 17~GHz PSR J1740--3015 appears to show a narrow, double peaked profile, with total $10\%$ width of $2.1^{\circ}$ (Figure \ref{J1740-3015_17000}).
We note that the 8.3~GHz profile shown in \citeN{jkw06} is in error, as it appears to have been broadened by an incorrect folding period, and therefore we include the corrected data in Figure \ref{J1740-3015_8356}.
As seen in Figure \ref{J1740-3015_17000}, the 17~GHz profile is very similar to that at 8.3~GHz, however the polarisation fraction has decreased, with linear fraction going from $\sim 80\%$ to $\sim 60\%$ and circular fraction from $\sim 60\%$ to $\sim 50\%$.
This is contrary to lower frequencies where the polarisation fraction is increasing with frequency \cite{gl98}.
At lower frequencies the profile appears to be somewhat scattered, but the 3.1~GHz profile appears to have two main components, with the leading component much brighter than the trailing, however the ratio of the two components appears to be the same between 8.3 and 17~GHz.


\section{Discussion}

\subsection{Polarisation fraction}
Of the pulsars for which the signal-to-noise ratio is great enough to reliably determine the polarisation fraction, there seems to be a division into almost $100\%$ polarised and essentially unpolarised.
Of particular note is PSR J1721--3532, for which there is significantly less linear polarisation at 17~GHz than at 8.3~GHz.
The MSP J0437--4715 appears unpolarised, however the polarisation fraction at low frequencies is small, and therefore we cannot be sure if the polarised fraction is different at 17~GHz.
The youngest pulsars in our sample, J0835--4510, J0908--4913, J1740--3015 and the magnetar J1622--4950 all retain a high fraction of linear polarisation at 17~GHz.

It should be noted that the literature is unclear weather a high linear polarisation fraction is predominantly associated with young age or high $\dot E$.
Most recently \citeN{wj08} showed that there is a clear division in $\dot E$ between low and high polarisation pulsars.
The radio magnetars are, however, clear exceptions to this rule since their long long spin period gives them a low $\dot E$ and yet they are highly polarised.
In any case, the highly polarised pulsars with high $\dot E$ and/or small age remain highly polarised throughout the observable frequency range.
In the framework of competing orthogonal polarisation modes (OPMs), this implies that the dominant OPM has a flatter spectral index than the weaker OPM (see e.g. Figure 2 of \citealp{kjm05}).
The only exception to this rule is PSR J0659+1414 which shows an abrupt decrease in polarisation between 3 and 5~GHz \cite{hoe99,jkw06}.
\citeN{jw06} and \citeN{kj07} have argued that these highly polarised pulsars also have high emission heights and so it appears that high polarisation fraction and high emission height may be lined in some way.
What remains even more difficult to explain in the context of emission models is the absence of polarisation over a broad frequency range.
In the OPM picture, this implies that the strengths of the modes are nearly equal, which requires fine tuning, especially considering that each OPM undergoes an independent exponential growth during its production \cite{mj04}.

\subsection{PSR J0835--4510}
\label{0835_model_section}

\begin{figure}
\includegraphics[height=8cm,angle=-90]{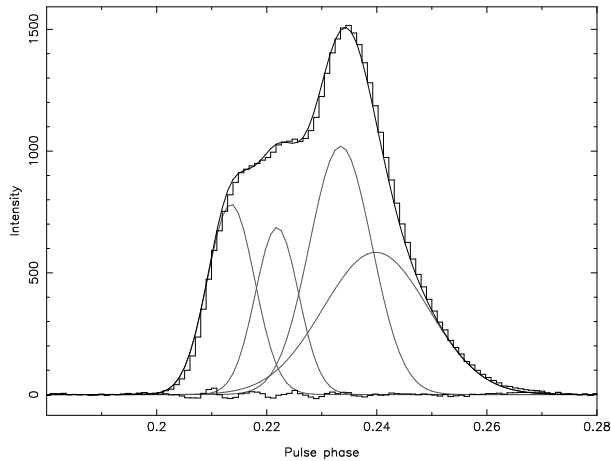}\caption{
        \label{0835_comp}
The four component model fit to total intensity profiles of J0835--4510.
Here the 6.3~GHz profile is shown, as this is where the four components are most equal in magnitude.
The upper histogram shows the 6.3~GHz profile of J0835--4510, whilst the solid line that traces it shows the model profile, composed of the sum of the four von Mises functions shown inside.
The lower histogram shows the residual data after subtraction of the model.
}\end{figure}

As seen in Figure \ref{J0835-4510_mf} the profile of PSR J0835--4510 has a fairly constant width between 1.4 and 24~GHz, however the relative amplitude of the various components varies with frequency.
Therefore we attempt to measure the spectrum of each component from 1.4 to 24~GHz.
In order to do this, we model the profile as a sum of a number of symmetric components represented by scaled \citeN{von18} functions, as implemented in the {\sc paas} software provided as part of {\sc psrchive}\footnote{http://psrchive.sourceforge.net}.
We find that four such components are required to model the observed profile, as shown in Figure \ref{0835_comp}.
Significantly, we found that we could fit the same model to all frequencies whilst keeping the width and separation of each component fixed.
Allowing the width and phase of the components to vary did not significantly improve the fit to the data, therefore we kept these properties fixed to the values given in Table \ref{0835_comp_tab}.

\begin{figure}
\includegraphics[width=8cm]{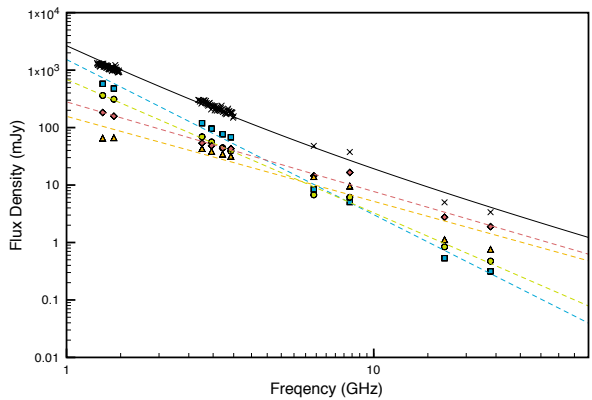}\caption{
        \label{0835_spec}
The flux density of PSR J0835--4510 as a function of frequency.
The small crosses represent the entire profile.
The model components are marked, in order of phase, by squares, circles, diamonds and triangles.
The dashed lines show spectral fits to the four components, with spectral indices as described in the text.
The solid line is the sum of the four dashed lines.
}\end{figure}

The flux densities of the each component as a function of frequency are presented in Figure \ref{0835_spec}, along with the observed flux density of the entire profile.
We fit a power law to the spectrum of each component, with flux $S = A \nu^{-\alpha}$, where $\alpha$ is the spectral index and $A$ is an arbitrary constant.
The best fit spectral indices are also given in Table \ref{0835_comp_tab}.
As can be seen by eye, the leading components have steeper spectra ($\alpha_A = 2.7$, $\alpha_B = 2.3$) and the trailing flatter ($\alpha_C\simeq\alpha_D=1.5$).

Therefore we conclude that the profile is made up of two pairs of components, the leading pair with a steep spectral index of $\sim 2.5$ and the trailing two with a shallower spectral index of $\sim 1.5$.
This supports the suggestion that the leading part of the profile is associated with `core' emission and the trailing edge is caused by a partial `cone' \cite{jvkb01,jkw06}.
The geometry of PSR J0835--4510 is well known, fitting to the observed position angles using the rotating vector model giving the magnetic inclination angle, $\alpha = 53^\circ$, and the smallest angle between the line of sight and the magnetic axis, $\beta = -6^\circ$ \cite{jhv+05}.
By considering the effects of aberration and retardation, \citeN{jvkb01} derive the emission height of the core to be only $\sim 100$~km.
What is less certain is the height of the conal emission.
\citeN{jvkb01} argue that the conal emission height is similar to that of the core, and that the cone that dominates at higher frequencies is an ``inner cone'', and that the so-called ``bump emission'' observed at the trailing edge of the pulse is the ``outer cone''.
For the inner cone we expect little evolution of the location relative to the magnetic pole with observing frequency \cite{mr02a}, consistent with what we derive for J0835--4510.
A plausible picture emerges therefore where all the emission over the frequency range 1 to 24~GHz is consistent with originating in the central or trailing portion of the open field lines at a height of 100~km, although a wider range of conal heights is not ruled out by the data.

It is clear from this analysis that the overall spectrum will flatten with frequency, changing from $\sim 2.5$ when dominated by the leading component to $\sim 1.5$ when dominated by the trailing component.
This suggests that, at least for this pulsar, the perceived change in spectral index with frequency is not intrinsic to the mechanism of the emission, but is caused by differing spectra between emission regions.
It is impossible for this effect to cause the spectral index to steepen at high frequencies, however it could potentially be the cause of the observed flattening of spectra in other pulsars \cite{kjdw97}.


\begin{table}
\begin{center}
\caption{\label{0835_comp_tab}
The phase ($\phi$) concentration ($\kappa$) and width ($W=360^{\circ}/\sqrt{\kappa}$) of the four von Mises functions used to model the profile of PSR J0835--4510. The amplitude of each component was kept free when fitting to the data.
The best fit spectral index is given in the last column.
}
\begin{tabular}{ccccr@{$\pm$}l}
Component & $\phi$ ($^\circ$) & $\kappa$ & $W$ ($^\circ$) &\multicolumn{2}{c}{$\alpha$}\\
\hline
A & -7.17 & 1384.7 & 9.67 & $2.7$&0.1\\
B & -4.17  & 1648.6 & 8.86 &$2.32$&0.06\\
C & 0 & 790.0 & 12.8       &$1.56$&0.06\\
D & 2.29 & 277.0 & 21.6    &$1.5$&0.2\\
\end{tabular}
\end{center}
\end{table}

\subsection{PSR J1622--4950}

\begin{figure}
\includegraphics[width=8cm]{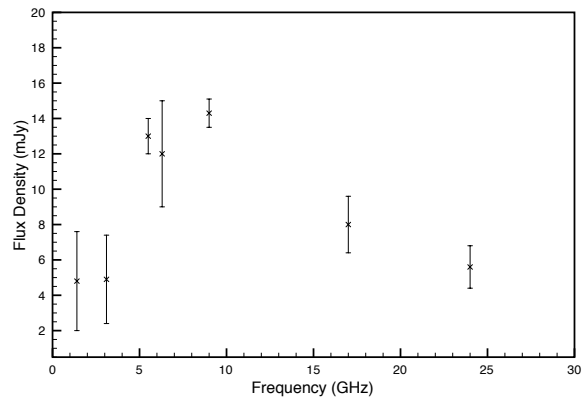}
\caption{
	\label{mag_flux}
The flux density of PSR J1622--4950 from 1.4 to 24 GHz.
}
\end{figure}

The radio magnetar PSR J1622--4950 has the highest 17~GHz flux density of any pulsar in our sample.
The flux density of PSR J1622--4950 as a function of frequency is plotted in Figure \ref{mag_flux}.
Data from 1.4 to 9~GHz are taken from \citeN{lbb+10}, the 17 and 24~GHz values are from this work.
The values at 1.4 and 3.1~GHz are statistical averages, whereas the values from higher frequencies are all single observations.
The intrinsic variability of this source makes it hard to be certain about the true nature of its spectral variation, however it is clear that the flux density varies by no more than a factor of 2 over a factor of nearly 20 in frequency.
Therefore, we conclude that the spectral index is very close to flat between 1.4 and 24~GHz.

\subsection{Future high frequency observations}
Of great interest in the pulsar and wider physics community is the
possibility of discovering pulsars in orbit about the black hole at
the Galactic Centre \cite{kbc+04}. Searches at low frequencies
have made advances on this \cite{jkl+06,dcl09} but the extreme
scattering at the Galactic Centre itself likely means that observing
frequencies well in excess of 10~GHz are required \cite{cl02} and
so far, searches at higher frequencies have proved fruitless \cite{mkfr10}.

Our observations, coupled with those of the Bonn group \cite{wjkg93,kjdw97} show that both normal and millisecond pulsars are detectable at 15~GHz and beyond and that we see only small changes in the spectral index from low frequencies to high. We therefore can extrapolate the known flux densities of pulsars from 1.4~GHz to higher frequencies to determine potential detectability at 15~GHz. \citeN{mkfr10} went through this exercise and showed that their lack of detections at 15~GHz implied a population of at most a few hundred pulsars in orbit around the Galactic Centre. We note that the magnetars as a class have long periods, spectral indices near zero and relatively high luminosities. These type of pulsars should be detectable in the Galactic Centre at high frequencies with current instrumentation if they exist there. For more typical pulsars, it seems that we must await the advent of the large collecting area of the Square Kilometre Array if we are to detect them at the Galactic Centre.

\section{Conclusion}
We have successfully detected 9 radio pulsars at a frequency of 17~GHz, including the highest ever detection of a millisecond pulsar, PSR J0437--4715.
Modelling the profile variation of PSR J0835--4510 as a function of frequency indicates that each component does not vary in width or phase, and has a well defined spectral index.
The difference between the spectral indices in each component cause the overall spectral index to flatten above $\sim 10$~GHz.

Additionally, we have detected the radio magnetar PSR J1622--4950 at a 17 and 24~GHz, noting that the profile looks almost identical to that at 3.1~GHz.
The flat spectrum and lack of frequency dependent profile shape changes firmly suggests that these properties are intrinsic to the class of radio-emitting magnetars.

\section{Acknowledgements}
The Parkes Observatory is part of the Australia Telescope which is funded by the Commonwealth of Australia for operation as a National Facility managed by CSIRO.

\bibliographystyle{mnras}
\bibliography{journals,myrefs,modrefs,psrrefs,crossrefs}

\end{document}